\documentstyle[a4,12pt]{article}

\def\journal#1#2#3#4{{#1} {\bf #2}, #3 (#4)}

\def\NPB{{\em Nucl. Phys.} B}
\def\PLB{{\em Phys. Lett.}  B}

\def\PRA{{\em Phys. Rev.} A}
\def\PRD{{\em Phys. Rev.} D}
\def\PRe{\em Phys. Rept.}

\newcommand{\Sslash}[1]{ \parbox[b]{0.6em}{$#1$} \hspace{-0.55em}
                         \parbox[b]{0.55em}{ \raisebox{-0.2ex}{$/$} }    }

\newcommand{\half}{\frac{1}{2}}
\newcommand{\gsim}{\buildrel > \over {_\sim}}
\newcommand{\lsim}{\buildrel < \over {_\sim}}
\newcommand{\ie}{{\it ie}}

\newcommand{\cf}{{\it cf}}

\newcommand{\ieps}{i\epsilon\,}
\newcommand{\tinf}{t=\pm\infty}
\newcommand{\limt}{{\buildrel {\scriptstyle{t_i\to -\infty}} \over {t_f\to
                    +\infty}}}

\newcommand{\pvec}{\vec p}
\newcommand{\kvec}{\vec k}

\newcommand{\ksla}{\Sslash{k}}
\newcommand{\lqcd}{\Lambda_{QCD}}
\newcommand{\ket}[1]{\vert{#1}\rangle}

\newcommand{\abs}[1]{\vert{#1}\,\vert}

\newcommand{\order}[1]{${\cal O}(#1)$}

\newcommand{\eq}[1]{Eq.\ (\ref{#1})}
\newcommand{\eqs}[2]{Eqs.\ (\ref{#1}),\ (\ref{#2})}

\def\be{\begin{equation}}
\def\ee{\end{equation}}
\def\bea{\begin{eqnarray}}
\def\eea{\end{eqnarray}}

\begin{document}

\begin{titlepage}
\begin{flushright}
        NORDITA--97/44 P\\
        hep-ph/9709444\\
       \today
\end{flushright}

\vskip 2.5cm

\centerline{\Large \bf A Perturbative Gluon Condensate\footnote{ Talk
presented at the `APCTP-ICTP Joint International Conference '97 (AIJIC 97)
on Recent Developments in Non-perturbative Methods', May 26 - 30, 1997 in
Seoul, Korea. Work supported in part by the EU/TMR contract ERB
FMRX-CT96-0008.}}

\vskip 1.5cm

\centerline{\bf Paul Hoyer}
\centerline{\sl Nordita}
\centerline{\sl Blegdamsvej 17, DK--2100 Copenhagen \O, Denmark}

\vskip 2cm

\begin{abstract} \noindent There is considerable freedom in setting boundary
conditions to perturbation theory at $\tinf$. The standard PQED and PQCD
expansions are based on the (empty) perturbative vacuum. Since the true QCD
ground state is expected to have a condensate of low momentum gluons, I
propose studying expansions around states containing free particles. A
specific boundary condition (the `Perturbative Gluon Condensate'), which
implies a freezing of low momentum gluons, could be a candidate for
describing confinement physics. The modified perturbative series has a
number of new features, although it is formally equivalent to the standard
one and {\em a priori} equally justified.
\end{abstract}

\end{titlepage}

\newpage
\renewcommand{\thefootnote}{\arabic{footnote}}
\setcounter{footnote}{0}
\setcounter{page}{1}

\section{Introduction}

\subsection{Perturbative Confinement?} \label{seconeone}

In this talk I would like to raise an unorthodox question \cite{pgc}:  {\em
Could light hadron structure be described by perturbative QCD methods?} A
perturbative expansion depends on the boundary conditions (\ie, states) at
asymptotic times $\tinf$. Thus, expanding around a state which includes low
momentum gluons gives a perturbation theory which at finite orders differs
importantly from the standard one, yet is formally equivalent to it. Both
expansions are {\em a priori} equally justified. The degree of freedom in
setting boundary conditions seems worth more attention than has been
accorded it so far.

The following facts seem to suggest that a perturbative expansion could be
relevant even for long-distance (confinement) physics in QCD:

\begin{itemize}
\item {\em Quarks and gluons are the relevant degrees of freedom in hadron
wave functions.} It is not always the case that the basic fields of the
lagrangian are directly reflected in physical quantities. Thus QED$_2$ may
be formulated either in terms of boson or fermion fields, and at strong
coupling $e/m \gg 1$ the bosonic formulation is more immediately related to
the bound state spectrum \cite{coleman}. In this respect QCD$_4$ is closer
to a weak coupling theory, suggesting the relevance of a perturbative
expansion.

\item {\em The semiquantitative success of the constituent quark model.} The
meson $(q\bar q)$ and baryon $(qqq)$ spectrum resembles that of QED atoms,
for which perturbation theory works well. We need to identify a systematic
approximation scheme for QCD in which the quark model is the first term.
\end{itemize}

In addition to raising the above question about a perturbative treatment of
long-distance QCD processes I shall propose specific boundary conditions at
$\tinf$. These conditions lead to a `freezing' of low-momentum gluons by
creating the analogue of a Fermi condensate for bosons. The corresponding
change in the Feynman rules only affects the `$\ieps$' prescription of
low-momentum gluon propagators.

\subsection{QED Perturbation Theory} \label{seconetwo}

QED Perturbation Theory is in amazing agreement with data --- particularly
in view of the fact that the expansion is asymptotic, with zero radius of
convergence. In the absence of experimental verification we would presumably
be uncertain whether PQED is relevant at all.

The standard PQED expansion is based on the perturbative vacuum
$\ket{\Omega_0}$ at $\tinf$, which has only an empty Fock state. The neglect
of fermion pairs $(e^+e^-,\ \mu^+\mu^-,$ $q\bar q,\ldots)$ and photons
$(\gamma)$ is a radical simplification of the exact (interactive) vacuum
$\ket{\Omega}$, which contains an infinity of particles.

Formally, a perturbative expansion around $\ket{\Omega_0}$ contains all
effects due to the exact vacuum which can be expanded in powers of $\alpha$.
In the infinite time available, any asymptotic ($\tinf$) state will `relax'
to the true lowest energy eigenstate  $\ket{E_0}=\ket{\Omega}$,
\be
\lim_{t \to \infty}\ket{\Omega_0,t}=
\lim_{t \to \infty}\sum_{n=0}^\infty c_n\,e^{-iE_nt}\ket{E_n} =
c_0\,e^{-iE_0t}\ket{E_0} \label{one}
\ee provided only that the overlap coefficient $c_0\neq 0$ (and that $t \to
\infty e^{-i\delta}$ with $\delta>0$). A perturbative description of
\eq{one} would require unlimited orders of the coupling constant $\alpha$.
The rapid convergence of the PQED series indicates that higher Fock states
of the true vacuum $\ket{\Omega}$ are short-lived and have a small
(perturbative) influence on physical observables.

\subsection{QCD: A Gluon Condensate} \label{seconethree}

The observed confinement of color shows that the QCD ground state has a
non-trivial long-distance ($\gsim 1/\lqcd \simeq 1$ fm) structure. In terms
of a Fock expansion based on the perturbative vacuum, this implies important
Fock components with low-momentum ($\abs{\pvec}
\lsim \lqcd$) gluons, forming a `gluon condensate' \cite{con}. It is then
not surprising that standard PQCD fails to describe long-distance physics
since the free gluon and quark propagators will (particularly at low
momenta) be strongly modified by the condensate gluons. Even though the
usual perturbative expansion is formally correct due to \eq{one}, the high
orders in $\alpha_s$ required to generate the condensate effect, together
with the asymptotic nature of the series, renders the expansion unreliable.

The question then arises whether we can find a perturbative series which
incorporates more of the long-distance QCD physics by expanding around an
asymptotic $(\tinf)$ state which contains low-momentum gluons. Such a
`Perturbative Gluon Condensate' \cite{pgc} could at best be only a crude
approximation of the exact QCD ground state -- but then an analogous
simplification (the perturbative vacuum) works well in QED. As discussed
above, the relatively simple hadronic structure suggested by the data and
the success of the quark model are central motivations for such an approach.

\section{Lorentz Invariance} \label{lorsect}

\subsection{Physical Measurables vs. Intermediate Quantities}
\label{sectwoone}

The exact QCD ground state is Lorentz invariant. It is indeed remarkable
that the gluon condensate, which is characterized by a momentum scale
$\abs{\pvec} \lsim \lqcd$, is invariant under arbitrarily large boosts. This
is possible only for an eigenstate of the {\em full} hamiltonian, for which
the frame dependence of the interactions can compensate that of the kinetic
part. Perturbative momentum states, which are eigenstates of the {\em free}
hamiltonian, are not invariant under Lorentz transformations. The only
exception is the perturbative vacuum, which has 4-momentum $P=0$. The fact
that any asymptotic state which contains free particles implies a
perturbative expansion which is not Lorentz-invariant order by order is
probably a major reason that an approach such as the one I am advocating
here has not been seriously pursued.

I shall argue that giving up Lorentz invariance at each order of
perturbation theory is neither incorrect nor unnatural in a theory like QCD
where all asymptotic states are bound states of the elementary fields. In
fact, even in QED bound state calculations it has been found useful to
consider non-covariant perturbative expansions \cite{lep}. The bound state
wave functions themselves need not be (and generally are not) Lorentz
covariant -- only physical (asymptotic) quantities like bound state masses
and scattering amplitudes are required to obey Lorentz symmetry. With both
the wave functions and the interactions between the elementary fields being
described by a perturbative expansion, it is a matter of convention which
interactions are regarded as belonging to the bound state and which to the
scattering subprocesses. The only crucial issue is that the combined object,
the S-matrix with hadronic external states, is correctly given by the theory
and therefore Lorentz invariant. This is indeed guaranteed by \eq{one} for
any perturbative expansion, where $\ket{\Omega_0}$ is an asymptotic state
having a non-vanishing overlap with the true ground state.

A specific example of how the interpretation of a physical process can
depend on the frame is provided by deep inelastic lepton scattering (DIS).
In the infinite momentum frame or light cone picture the photon strikes a
quark carrying momentum fraction $x_{Bj}$. The cross section is proportional
to the probability of finding such a quark in the target, as given by the
target structure function. In the target rest frame, on the other hand, the
same quark appears as part of the photon wave function, and the DIS cross
section is determined by the probability for the quark to interact with the
target. Both descriptions must lead to the same physical cross section,
implying a relation between the quark structure function of the target and
the quark-target cross section. It is not possible to explicitly demonstrate
this relation using PQCD since it involves the (non-perturbative) proton
wave function in different frames.

\subsection{Light Cone vs. Equal Time Quantization} \label{sectwotwo}

The Light Cone (LC) and Equal Time (ET) quantizationn schemes provide an
example of two formalisms which have different Lorentz properties but which
must give equivalent results for physical quantities. In LC quantization,
explicit boost invariance is preserved along the LC, whereas rotational
invariance is lost. This is an advantage in calculating transition form
factors, which involve boosted bound states. In particular, amplitudes for
hard QCD processes are most simply expressed in terms of LC wave functions
\cite{brle}.

The LC formalism is less attractive for bound state wave functions in the
rest frame, where physics is rotationally invariant. Gluon condensate
effects emerge indirectly in the LC framework. The perturbative vacuum is an
eigenstate of the full LC hamiltonian, and thus should be the true ground
state of the theory. Condensate effects are understood as originating from
`zero modes', which arise because the quantization surface connects points
at light-like distances which thus are causally connected \cite{pinsky}.

In order to keep the physically intuitive notion of a gluon condensate and
to maintain rotational invariance I shall use equal time quantization. In
the ET framework there is no simple relation between hard QCD processes in
the `infinite momentum frame' and hadronic wave functions in the rest frame.
This may actually not be a disadvantage, since hadron structure appears
empirically to be frame-dependent. Thus gluons carry about half the hadron
momentum in the infinite momentum frame, but appear frozen into the
constituent quark mass in the rest frame.

\subsection{Frame Dependent Perturbation Theory} \label{sectwothree}

Consider, then, that we are expanding around a boundary state at $\tinf$
which contains free gluons with momenta $\abs{\pvec} \leq \lqcd$. Since this
is our model for the true gluon condensate, we have to use the {\em same}
state in all frames. Thus a calculation of a given Green function in two
different frames will generate distinct perturbative expansions, since the
boundary condition is not boosted. The expansions will be formally
equivalent, but to explicitly demonstrate how they are related is no easier
than demonstrating \eq{one}, or showing the boost invariance of the true
ground state. Physical quantities (masses, cross sections, etc.) should
nevertheless turn out to be the same, within the accuracy of the
calculation. This situation is somewhat analogous to the well-known
dependence on the renormalization scale $\mu$ in standard PQCD calculations.
According to the (non-perturbative) renormalization group we know that the
exact answer is independent of $\mu$, but to any finite order in
perturbation theory the result is $\mu$-dependent. One is then forced to use
other arguments to decide which value of $\mu$ gives optimal results.
Similarly, when using the same boundary condition in all frames one will
have to consider which frame is best suited for the calculation of a given
quantity.

In simple cases where the exact result is known one can verify explicitly the
boost invariance of physical quantities. That a frame dependent equal-time
bound state equation can generate a covariant four-momentum is illustrated
by the following 1+1 dimensional example \cite{boo}. The wave function of a
two fermion bound state is written
\be
\psi(t,x_1,x_2) = \exp(-iEt)\, \exp\left(ik\frac{x_1+x_2}{2}\right)\,
\chi(x_1-x_2)~~,  \label{wf}
\ee where $x_1,x_2$ are the positions of the constituents and $t$ their
common time. Both the bound state energy $E$ and the $2\times 2$ Dirac wave
function $\chi$ depend on the bound state c.m. momentum parameter $k$. The
bound state equation for $\chi$ is then\footnote{I expect this equation to
be accurate for QED$_2$ in the limit of weak coupling, $e/m \ll 1$, \ie, for
non-relativistic internal motion, but for arbitrary values of the center of
mass momentum $k$.}
\be -i\partial_x \left[\alpha,\chi(x)\right] +\half k
\left\{\alpha,\chi(x)\right\} +m_1\gamma^0 \chi(x) -m_2\chi(x) \gamma^0 =
\left(E-V(x)\right) \chi(x)  \label{bse}
\ee where $\alpha=\gamma^0\gamma^1$,  $m_{1,2}$ are the constituent masses
and
$V(x)=\half e^2 |x|$ is the instantaneous Coulomb potential of QED$_2$.
Despite the fact that
\eq{bse} has no explicit Lorentz covariance (space and time coordinates are
treated differently in \eqs{wf}{bse}) the bound state energies for different
c.m. momenta $k$ are correctly related:
$E=\sqrt{k^2+M^2}$, with
$M$ independent of $k$. In the limit of non-relativistic internal motion
$(e/m_{1,2} \ll 1)$ the wave function $\chi(x)$  Lorentz contracts in the
standard way as a function of $k$. Interestingly, the hidden Lorentz
covariance of \eq{bse} holds only in the case of a linear potential $V(x)$.

\section{Filling the Perturbative Vacuum} \label{secthree}

\subsection{The Free Fermion Propagator} \label{secthreeone}

In considering the effects on perturbation theory of including particles in
the asymptotic states at $\tinf$ it is simplest to begin with fermions. Due
to the Pauli principle we can have only 0 or 1 fermions in a state of given
3-momentum and spin. As we shall see, this exercise will also suggest an
interesting boundary condition for bosons.

The standard free fermion propagator
\be iS_F(x-y)= \langle 0|T[\psi(x)\bar\psi(y)]|0\rangle   \label{sfxy}
\ee is in momentum space
\be S_F(p)= \frac{\Sslash{p}+m}{p^2-m^2+\ieps} =
\frac{\Sslash{p}+m}{(p^0-E_p+\ieps)(p^0+E_p-\ieps)}~~.  \label{sfp}
\ee If we add an antifermion to the initial and final states,
\bea
\langle 0|d_{\lambda'}(\kvec')T[\psi(x)\bar\psi(y)]d_{\lambda}^{\dag}
(\kvec)|0\rangle &=& iS_F(x-y)2E_k (2\pi)^3
\delta^3(\kvec-\kvec')\delta_{\lambda\lambda'}
 \nonumber \\ &+& v(\lambda',\kvec') \bar v(\lambda,\kvec) e^{ik'\cdot x -
ik\cdot y}~~,
\label{onef}
\eea there is a new term corresponding to a mixing of the antifermion
propagating from $x$ to $y$ with the antifermion in the in- and out-states.

For a condensate we would fill both helicity states at a given momentum
$\kvec$. The free propagator
\be iS(x-y)\equiv \langle 0|d_\half(\kvec)d_{-\half}(\kvec)T[\psi(x)\bar
\psi(y)]d_{-\half}^{\dag}(\kvec) d_\half^{\dag} (\kvec)|0\rangle
\label{twofxy} \ee is then in momentum space
\be S(p)= \left[(2\pi)^3 2E_k \delta^3(\vec 0)\right]^2 \left\{
\begin{array}{ll}S_F(p)&\ \ \ \ (\pvec \neq -\kvec) \\
                 S_E(p)&\ \ \ \ (\pvec = -\kvec) \\
\end{array} \right.  \label{twofp}
\ee where
\be S_E(p)= \frac{\Sslash{p}+m}{(p^0-E_k+\ieps)(p^0+E_k+\ieps)}  \label{sep}
\ee differs from the Feynman propagator only in the $\ieps$ prescription at
$p^0=-E_k$. Since the antifermions inserted in the definition (\ref{twofxy})
are on-shell, it is clear that a mixing between them and the propagating
fermion only can occur at the antifermion pole of $S(p)$. Adding
antifermions for all momenta $\abs{\kvec} \leq \Lambda$, the corresponding
propagator will equal $S_E(p)$ for all $\abs{\pvec} \leq
\Lambda$.

\subsection{Fermion Generating Functionals}

The above analysis for the free propagator carries over to arbitrary Green
functions at any order of perturbation theory. To see this it is best to use
the generating functional of Green functions, which in a typical gauge
theory is of the form

\bea Z[J;\zeta,\bar\zeta] &=& \int{\cal D}[A] {\cal D}[\bar\psi,\psi]
\exp\left[\int d^4x (i {\cal L}+iJ\cdot A + \bar\zeta\psi + \bar\psi\zeta)
\right] \nonumber \\ &=& \exp\left[ iS_{int}\left(\frac{\delta}{\delta
iJ};\frac{-\delta}{\delta\zeta},
\frac{\delta}{\delta\bar\zeta}\right)\right] Z[J]Z[\zeta,\bar\zeta]
\label{genfun}
\eea where $S_{int}$ is the interaction part of the action and
$Z[J],~Z[\zeta,\bar\zeta]$ are free functionals of the boson $(J)$ and
fermion $(\zeta,~\bar\zeta)$ sources, respectively. The free fermion
functional is
\be Z_{F,E}[\zeta,\bar\zeta]=\exp\left[i\int\frac{d^4p}{(2\pi)^4}
\bar\zeta(-p) S_{F,E}(p) \zeta(p) \right]  \label{fgenfun}
\ee with the fermion propagators $S_{F,E}$ given by \eqs{sfp}{sep}. We can
relate the free generating functionals $Z_E,\ Z_F$ to each other using
\be S_E(p)=S_F(p)+2\pi i\delta(p^0-k^0)
\frac{\delta^3(\pvec-\kvec)}{\delta^3(0)}
\frac{\ksla+m}{2E_k} \label{sefrel}
\ee where $k^0=-E_k=-\sqrt{\kvec^2+m^2}$. Hence,
\bea Z_E &=& \exp\left[\frac{1}{(2\pi)^3\delta^3(0)2E_k}\bar\zeta(-k)
\sum_\lambda v(-\kvec,\lambda) \bar v(-\kvec,\lambda)\zeta(k)\right] Z_F
\label{zefrela} \\ &=& \prod_\lambda \frac{1}{(2\pi)^3\delta^3(0)2E_k} \left[
\frac{\delta}{\delta\zeta}S_F^{-1}  v(-\kvec,\lambda)\right] \left[ \bar
v(-\kvec,\lambda) S_F^{-1} \frac{\delta}{\delta\bar\zeta}\right] Z_F~~.
\label{zefrelb}
\eea In \eq{zefrela} the expansion of the Grassman valued exponential gives
two terms (for each helicity $\lambda$), which correspond to the two terms of
\eq{zefrelb} where the source derivatives operate on the same or different
exponents in $Z_F$.

Since the source derivatives in \eq{zefrelb} correspond to inserting
antifermions in the initial and final states, this relation generalizes our
previous result of \eqs{twofxy}{twofp} for the free propagator to that of
the free generating functional (\ref{fgenfun}). Now relation (\ref{zefrelb})
can be immediately carried over to the full generating functional
(\ref{genfun}) of interactive Green functions since the source derivatives
in \eq{zefrelb} commute through the source derivatives in $S_{int}$. We have
thus reached the conclusion that

{\em Any Green function (to arbitrary order in the coupling $\alpha$)
evaluated with $S_E$ propagators gives the same result as a standard
calculation using $S_F$ if antifermions are added in the initial and final
states.}

Thus we can perform perturbative calculations in the presence of a Fermi
condensate, with all momentum and spin states occupied up to a Fermi
momentum $p_f$, by changing the $\ieps$ prescription of the free propagator
below $p_f$. The standard causality argument for the sign of $\ieps$ in the
propagator is not applicable since we are dealing with interference between
several particles.

Asymptotic QCD states with extra (anti-)quarks in the initial and final
states have non-vanishing baryon number and hence no overlap with the true
vacuum (\cf\ \eq{one}). It is interesting to ask, however, whether a
statement analogous to the one above can be made for gluons (\ie, for
bosons). Clearly we can change (`by hand') the sign of
$\ieps$ also in boson propagators, but is that equivalent to adding bosons to
the initial and final states? Moreover, our fermion example is in a sense
similar to the confinement situation in QCD. With all fermion states
occupied up to the Fermi momentum $p_f$, no fermion-antifermion pair
production can occur for $\abs{\pvec}\le p_f$. The
$\ieps$ sign change in the fermion propagator accomplishes this by preventing
a pinch between poles on different sides of the real axis. This will also
happen if the sign of $\ieps$ is changed in boson propagators, and would thus
prevent the production of low momentum (confined) gluons.

\subsection{Boson Generating Functionals}

The generating functionals for free scalar bosons are given by
\be Z_{F,E}[J]= \exp\left[\frac{i}{2}\sum_{\pm\pvec}\int\frac{dp^0}{2\pi}
J(-p^0,-\pvec) D_{F,E}(p) J(p^0,\pvec) \right]  \label{bgf}
\ee Since we shall be dealing with the free functional (the generalization to
the interacting one will again be straightforward), it is sufficient to
consider a single 3-momentum $\pvec$, and keep only the Bose symmetrization
over $\pm\pvec$ as indicated in \eq{bgf}. The Feynman propagator is
\be D_F(p)= \frac{1}{p^2-m^2+\ieps} = \frac{1}{2E}\left(
\frac{1}{p^0-E+\ieps}-\frac{1}{p^0+E-\ieps} \right)  \label{bfp}
\ee where $E=\sqrt{\pvec^2+m^2}$. A modification of the $\ieps$ prescription
at the $p^0=-E$ pole gives
\be D_E(p) \equiv \frac{1}{(p^0-E+\ieps)(p^0+E+\ieps)} = D_F(p)+ \frac{2\pi
i}{2E}\delta(p^0+E)  \label{bep}
\ee Note that the same generating functional is obtained if the $\ieps$
prescription is changed instead at the $p^0=+E$ pole, due to its symmetry
under $p \to -p$.

Substituting the relation (\ref{bep}) between the $E$ and $F$ propagators
into the expression (\ref{bgf}) for the generating functional $Z_E$ we find,
after a Fourier transform to $(t,\pvec)$ space,
\be Z_E[J] = \exp\left[ -\sum_{\pm\pvec}\frac{1}{4E}\int dt'dt''
J(t'',-\pvec) e^{iE(t''-t')}J(t',\pvec) \right] Z_F[J]  \label{ftbegf}
\ee

\eq{ftbegf} may be compared with \eq{zefrela} in the fermion case. Due to the
Grassmann algebra the exponential factor multiplying $Z_F[\zeta,\bar\zeta]$
contains only a single power of the fermion sources. In the boson case the
factor multiplying $Z_F[J]$ in \eq{ftbegf} contains arbitrary powers of the
source $J$. This factor can then be reproduced only by differentiating
$Z_F[J]$ an arbitrary number of times, corresponding to an indefinite number
of incoming and outgoing bosons. A straightforward calculation gives
\cite{pgc}
\be Z_E[J]=4\exp\left[-\sum_{\pm\pvec}\lim_\limt e^{iE(t_f-t_i)} 2E
\frac{\delta^2} {\delta J(t_f,\pvec)\delta J(t_i,-\pvec)} \right] Z_F[J]
\label{zeres}
\ee

The source derivatives in (\ref{zeres}) commute through the interaction term
in the definition (\ref{genfun}) of the full generating functional of Green
functions in the interacting theory. Hence the above result establishes that

{\em A change in the $\ieps$ prescription of the scalar propagator at a
given 3-momentum is equivalent (for any Green function and to arbitrary order
in $\alpha$) to a superposition of standard Green functions with
$0,1,2,\ldots$ scalars of that 3-momentum in the initial and final states.}

The situation is thus slightly different from the fermion case since we have
a superposition of Green functions. Nevertheless, each one of them will
generate an equivalent perturbative series according to the argument related
to \eq{one}, and therefore also their superposition gives a series formally
equivalent to the standard one\footnote{I am assuming a non-vanishing
overlap with the true vacuum. Such an assumption is tacitly made also for
the standard series based on the perturbative vacuum}.

\subsection{The Perturbative Gluon Condensate}

The above considerations suggest a specific boundary condition to QCD
perturbation theory, the {\em Perturbative Gluon Condensate,} which is
equivalent to changing the $\ieps$ prescription for gluon propagators with
$\abs{\pvec} < \lqcd$. It corresponds to a superposition of standard PQCD
series with a varying number of gluons of
$\abs{\pvec} < \lqcd$ included in the initial and final states. The
perturbative series with this boundary condition is {\em a priori} as well
justified as the standard one with no gluons added. Since none of the series
converges the equivalence between them is formal. Results based on low order
terms from different series can be quite distinct.

The consequences of this (superficially simple) change in the Feynman rules
of QCD need to be further explored in order to establish its possible
relevance for describing long-distance physics.  The following general
remarks can be made.

\begin{itemize}
\item Gauge symmetry is preserved, since only physical (on-shell transverse)
gluons are added to the initial and final states\footnote{Whether the
$\ieps$ prescription is changed for unphysical components of internal gluon
and ghost propagators should not matter, since those components do not
contribute to imaginary parts of the amplitude.}.

\item Lorentz invariance is formally valid for the full series, but broken
at any finite order. Only physical measurables, not given orders of
perturbation theory, are expected to be covariant when the asymptotic states
are not elementary. A loss of explicit boost invariance is expected in any
framework for equal-time bound states.

\item A change in the $\ieps$ prescription of gluon propagators will alter
the singularity structure and the unitarity relations for Green functions.
However, standard analyticity and unitarity should only be expected to apply
for physical amplitudes, having hadrons rather than quarks and gluons as
external particles. As has been explicitly demonstrated \cite{qcd2} in
QCD$_2$, Green functions involving confined external fields can have new
types of singularities even though physical amplitudes have normal
analyticity.

\item The standard argument based on gauge and Lorentz invariance that the
gauge fields remain massless to all orders in perturbatoin theory fails, due
to the absence of explicit Lorentz invariance. Thus an effective
(momentum-dependent) gluon mass is generated by loop diagrams. Similarly, it
may be expected that an effective quark mass is generated.

\item By construction, the gluon degrees of freedom are frozen for
$\abs{\pvec} < \lqcd$, in the same sense as for fermions below a Fermi
momentum. The perturbative gluon condensate creates the rather unusual
situation of a `Fermi sea for gluons', which at least superficially seems to
correspond to the observed absence of low-momentum gluons.

\item Since only low momentum gluon propagators are modified, the standard
PQCD results for hard QCD processes in the infinite momentum frame are
preserved. Similarly, the propagator modification has no effect on the
renormalization procedure, which concerns ultraviolet singularities.

\end{itemize}

\subsection{The Schwinger Model}

The general arguments of section 1 that equivalent perturbative series may
be obtained by expanding around different asymptotic states can presumably be
checked in exactly solvable models. A particularly simple case is the
Schwinger model \cite{coleman,ph} (QED$_2$ with $m_e=0$).

The only nonvanishing (1PI) Feynman diagram of the Schwinger model is the
fermion loop with two external photon of momentum $q$, which is
\be L_2^{\mu\nu}=-\frac{ie^2}{\pi} \left(-g^{\mu\nu}+q^\mu q^\nu/q^2 \right)
\label{l2}
\ee When iterated this diagram generates the Schwinger boson mass
$M^2=e^2/\pi$. The boson is non-interacting since fermion loops $L_n$ with
$n>2$ external photons vanish identically.

It is straightforward to establish that a change in the $\ieps$ prescription
of the electron propagator $S_F(p)$ at any given momentum $k^1$,
\be
\delta_k S_F(p) = \left[S_E(p)-S_F(p)\right] 2\pi\delta(p^1-k^1) =i
\frac{\ksla}{2\abs{k^1}}(2\pi)^2 \delta^2(p-k)  \label{dksf}
\ee leaves the values of the fermion loop diagrams unchanged. In particular,
all dependence on the momentum $k^1$ is cancelled and Lorentz (as well as
gauge) symmetry is preserved at each order of perturbation theory.

Using the $S_E(p)$ propagator for all momenta $\abs{p^1}< \Lambda$ there is
no contribution to $L_n$ from loop momenta below the arbitrary scale of
\order{\Lambda}. The fact that Schwinger model physics is independent of
$\Lambda$ reflects the pointlike nature of the Schwinger boson (which in the
bosonic formulation is described by an elementary, non-interacting scalar
field).

\section{Summary and Conclusions}

I have emphasized the possibility and interest of constructing formally
equivalent perturbative series by expanding around asymptotic states that
contain different numbers of free (`background') particles. The standard
series based on the (empty) perturbative vacuum is thus only one of many
alternatives. It seems worthwhile to investigate relations between the
series based on their formal equivalence.

For practical applications based on low order results it is clearly
important to expand around a state which is as good an approximation of the
true ground state as possible. The success of PQED shows that the
perturbative vacuum is, in spite of its simplicity, a good model for the
true vacuum of QED.

Conversely, the failure of PQCD to correctly describe long-distance hadron
physics implies that the true QCD vacuum contains a `gluon condensate',
characterized by a length scale $\lqcd^{-1} \simeq 1$ fm. It is natural to
ask whether a series based on an asymptotic state which contains
low-momentum gluons could give a perturbative description of confinement
physics. The simple yet qualitatively successful description of hadrons
provided by the constituent quark model lends empirical support to such an
effort.

The phenomenology of strong interaction physics suggests that the degrees of
freedom corresponding to low momentum $(\abs{\pvec} \lsim \lqcd)$ gluons are
frozen. This feature can be realized in perturbation theory through an
analogy with a Fermi condensate, which prevents particle production below the
Fermi momentum. As a matter of fact, perturbation theory in the presence of a
Fermi condensate is an example of an expansion around a non-empty asymptotic
state at $\tinf$. The condensate affects the Feynman rules only through a
change in the $\ieps$ prescription of the free fermion propagator.

A corresponding change of the $\ieps$ prescription for boson propagators
similarly removes bosonic degrees of freedom, and is again equivalent to a
change in the boundary conditions. In the bosonic case the modified Green
function corresponds to a superposition of standard Green functions with
0,1,2,\ldots\ bosons added to the initial and final states. Although the
$\ieps$-modified perturbative expansion is thus formally as well motivated as
the standard one, it has many novel features and it is far from clear whether
such an approach will prove useful in practice.

Perhaps the most startling feature of a perturbative series involving an
explicit momentum scale is that each order of the expansion is not
separately Lorentz invariant, although the full series does obey the
symmetry. This is expected for amplitudes involving bound asymptotic states,
since the bound state wave function generally does not have simple
properties under Lorentz transformations. Only physical measurables such as
invariant masses and cross sections should be frame-independent. Due to the
infinite powers of the coupling constant contained in the bound state wave
function this physical requirement does not imply that each order of
perturbation theory is manifestly invariant. In fact, QCD physics does
empirically depend on the frame. The constituent quark model picture of
(rest frame) hadrons is quite different from that observed in hard processes
(which correspond to the infinite momentum frame).

I am not aware of other approaches to QCD confinement physics that would be
closely related to what I have presented here. Gribov \cite{gribov} has
previously considered changes in the analytic structure of Green functions
due to strong interaction effects. More recently, a modified
$\ieps$ prescription of photon propagators emerged \cite{asz} in a study of
the KLN theorem.

\subsection*{Acknowledgement}

I am very grateful to the organizers of this meeting for their invitation
and hospitality.


\end{document}